\documentclass{article}
\usepackage{pdfpages}
\usepackage{xcolor}
\usepackage{graphicx}
\usepackage{cite}
\usepackage{caption}
\usepackage{subcaption}
\usepackage{authblk}
\usepackage{amsmath}

\usepackage[top=5cm, bottom=5cm, left=3cm, right=2cm, headsep=1.5cm, heightrounded=true]{geometry}
\usepackage{fancyhdr}
\graphicspath{ {{images/}} }

\title{Holographic Thermodynamics of Higher-Dimensional AdS Black Holes with CFT Rescaling  }

\author[1,2]{Yahya Ladghami\thanks{ \texttt{yahya.ladghami@ump.ac.ma}}}
\author[1,2]{Taoufik Ouali\thanks{ \texttt{t.ouali@ump.ac.ma}}}
\affil[1] {Laboratory of Physics of Matter and Radiation, Mohammed I University, BP 717, Oujda, Morocco}
\affil[2]{Astrophysical and Cosmological Center, BP 717, Oujda, Morocco}

\begin{document}
	\maketitle
	\begin{abstract}
		In this paper, we study the thermodynamic behavior of  charged AdS black holes in higher-dimensional spacetimes within the framework of conformal holographic extended thermodynamics. This formalism is based on a novel AdS/CFT dictionary in which the conformal rescaling factor of the boundary conformal field theory (CFT) is treated as a thermodynamic parameter, while Newton’s constant is held fixed and the AdS radius is allowed to vary. We explore how variations in the CFT state, represented by its central charge, influence the bulk thermodynamics, phase structure, and stability of black holes in five and six dimensions. Our analysis reveals the emergence of Van der Waals–like phase transitions, critical phenomena governed by the central charge.   Additionally, we find that the thermodynamic behavior of AdS black holes is affected by the dimensionality of the bulk spacetime, as we compare higher-dimensional black holes to lower-dimensional ones, such as the BTZ black holes.    These findings provide new insights into the role of boundary degrees of freedom in shaping the thermodynamics of gravitational systems via holography.

	\end{abstract}
	
	\section{Introduction}
	Black holes are considered as a crucial key to understanding the nature of spacetime and quantum gravity, as well as to addressing deep problems in theoretical physics, such as the information paradox and the unification of fundamental forces. Black holes gained significant attention following the pioneering works of Hawking and Bekenstein. Hawking discovered that black holes emit black-body radiation and evaporate due to quantum effects near the event horizon~\cite{1}. Bekenstein proposed that black holes possess an entropy proportional to the area of their event horizon, in contrast to ordinary thermodynamic systems where entropy typically scales with volume. This entropy is now known in the literature as the Bekenstein–Hawking entropy~\cite{2}. These results demonstrate that black holes are not merely gravitational objects, but also thermodynamic systems governed by four laws analogous to the laws of thermodynamics.~\cite{3}.
	\\
	
	Black holes in asymptotically Anti-de Sitter (AdS) spacetime, known as AdS black holes, form a cornerstone in the development of black hole physics. In this context, the cosmological constant is interpreted as a thermodynamic pressure and treated as a thermodynamic variable, while the mass of the black hole is identified with enthalpy~\cite{4}. This framework is referred to as \emph{extended phase space thermodynamics} (EPST), or alternatively as the thermodynamics of \emph{chemical black holes}~\cite{5}. Within this thermodynamic, several interesting phenomena of black holes have been discovered, such as the Van der Waals–like behavior of black holes~\cite{6,7,7a,7b,7c,8}, the Joule–Thomson expansion~\cite{9,10,11,12,13}, and the existence of multi-critical points~\cite{14,15,16}.
	\\
	
	The Anti–de Sitter/Conformal Field Theory (AdS/CFT) correspondence~\cite{17} provides a powerful framework for studying and understanding black hole physics. Many complex issues related to black holes have been investigated using AdS/CFT correspondence, including the information loss paradox~\cite{a17}, the island prescription for black hole entanglement~\cite{b17}, and holographic analyses of black-hole thermodynamics~\cite{18}. In this work we focus exclusively on the thermodynamics of black holes within the AdS/CFT correspondence. Furthermore, within the AdS/CFT  correspondence,  the thermodynamics of AdS black holes is interpreted as equivalent to the thermodynamics of the dual CFT, with the physical quantities on both sides related through the AdS/CFT dictionary. Using this correspondence, a first law of holographic thermodynamics has been formulated, involving a new thermodynamic pair \((\mu, C)\), where \(\mu\) is the chemical potential and \(C\) is the central charge of  CFT \cite{18}. However, this formulation suffers from degeneracy because the variations of the central charge and  CFT volume are not independent. To resolve this issue, two main approaches have been proposed.
	\\
	
	The first approach treats Newton’s constant as a variable while keeping the cosmological constant fixed~\cite{19}. This framework is commonly known in the literature as \emph{restricted phase space thermodynamics}  (RPST)~\cite{20,21,22,23,24,25}. The second approach, which we adopt in this work, keeps Newton’s constant fixed and allows the cosmological constant to vary. In this case, the conformal field theory is rescaled by a conformal factor, and treated as a thermodynamic parameter~\cite{26,27}. This thermodynamic framework is known as \emph{conformal holographic extended thermodynamics} (CHET) \cite{28}. Several classes of AdS black holes have been investigated within this framework, including Reissner–Nordström AdS black holes~\cite{28}, Gauss–Bonnet black holes~\cite{29}, and BTZ black holes~\cite{30}. The success of CHET in lower and four dimensions, as demonstrated in the aforementioned studies, motivates its extension to higher-dimensional cases.
	\\
	
	The present work extends our earlier studies~\cite{28,30} of black-hole thermodynamics with CFT rescaling to the case $D>4$. We are particularly interested in $D>4$ bulk gravity and the corresponding boundary CFT, focusing on higher-dimensional AdS black holes because they provide an excellent laboratory for probing the nature and microscopic description of spacetime and quantum gravity. Higher-dimensional black holes play a central role in many theoretical frameworks that address quantum gravity, since those theories naturally involve higher-dimensional spacetimes and describe systems that combine quantum and gravitational effects. The literature on higher-dimensional black holes is extensive, covering topics such as thermodynamic behaviour~\cite{hd}, geometric approaches to black-hole thermodynamics~\cite{hd1}, stability analyses of higher-dimensional Schwarzschild solutions~\cite{hd2}, quasinormal-mode spectra~\cite{hd3}, higher-dimensional solutions in modified gravity~\cite{hd4}, and topological thermodynamics~\cite{hd5}.
	\\

	Although our formalism is indeed extendable to arbitrary $D$, the analytic form of the thermodynamic quantities and critical parameters becomes increasingly complex for general $D$. For this reason, we focus on the explicit and physically motivated cases $D = 5$ and $D = 6$, where the phase structures can be analyzed in detail and compared directly to known results in four dimensions and BTZ black holes. Furthermore, the choice $D = 5$ and $D = 6$ is justified by their richer holographic and gravitational structures that emerge in higher dimensions. In particular, in $D=5$ and $D=6$, AdS black holes play a central role in string theory and higher-dimensional models of quantum gravity. They also provide an important arena to test whether the conformal holographic extended thermodynamics formalism exhibits universal properties across dimensions.
	\\
	
	Our primary goal is to highlight the physical relevance of the CFT at the boundary in dictating the behavior of gravitational systems in the bulk, within the context of the AdS/CFT correspondence. We study how variations in the boundary state, particularly through the central charge of the CFT, influence the bulk thermodynamics, with a focus on higher-dimensional black holes. Specifically, we analyze the impact of the CFT degrees of freedom on the thermodynamic behavior of black holes, including their stability, critical phenomena, and phase structure. These findings not only extend the scope of CHET formalism but also deepen our understanding of how the microscopic structure of the boundary theory governs macroscopic gravitational thermodynamics.
	\\

	This paper is organized as follows: In Section~\ref{SS1}, we present the framework of conformal holographic extended thermodynamics for black holes. In Section~\ref{SS2}, we investigate higher-dimensional charged AdS black holes,  by focusing on 5 and 6 dimensions, and analyze their thermodynamic behavior. In Section~\ref{SS3}, we summarize our findings and present concluding remarks. Throughout this paper, we adopt natural units where \(\hbar = c = k_B = 1\).
	\section{Conformal Holographic Extended Thermodynamic  }
	\label{SS1}
	In this section, we present  the holographic interpretation of AdS black holes within the context of the AdS/CFT correspondence~\cite{17}. In this framework, the thermodynamics of AdS black holes is understood to be equivalent to the thermodynamics of a conformal field theory (CFT). The thermodynamic quantities of the AdS black hole are related to those of the CFT via what is known as the AdS/CFT dictionary~\cite{18}.

	This approach serves as an alternative to other thermodynamic frameworks such as extended phase space thermodynamics~\cite{4}, Visser’s thermodynamics~\cite{18}, and restricted phase space thermodynamics~\cite{19}. A key advantage of this method is that it resolves the degeneracy issue\footnote{In the holographic first law, the variations of the CFT volume and the central charge are not independent~\cite{26}.} that arises in other holographic thermodynamic formalisms, without the need to treat Newton’s constant as a thermodynamic variable, as is done in restricted phase space thermodynamics.
	
	The cornerstone of this formalism lies in the rescaling of the CFT metric by a conformal factor, which is treated as a thermodynamic parameter. This setup allows for the establishment of a new AdS/CFT dictionary and a novel non-degenerate first law of holographic thermodynamics. In this framework, the bulk AdS radius differs from the radius of  CFT.
	
	The following metric illustrates the rescaling of  CFT \cite{31,32}
	\begin{equation}
		d s^2 = \omega^2 \left( -dt^2 + L^2 d \Omega^2_{d-2}\right), 
	\end{equation}
	where $\omega$ is a  dimensionless conformal factor, $L$ is the AdS radius, and $d$ represents the dimensionality of the bulk spacetime. As a result, the volume and time in the CFT are rescaled, respectively, as follows \cite{26}
	\begin{equation}
		\label{9Z}
		V \propto \left( \omega L\right)^{d-2}, \qquad \tilde{t} = \omega t
	\end{equation} 
	where $t$ is the bulk time. Trough  CFT time we find the energy of CFT, $\tilde{E}$, as follow \cite{28}
	\begin{equation}
		\tilde{E} =  \dfrac{d t}{ d \tilde{t}}\, E= \frac{E}{\omega}
	\end{equation} 
	where \(E\) is the AdS black hole energy and is equivalent to the black hole mass. Based on this setup, we derive the AdS/CFT dictionary as follows \cite{26} (the quantities with tildes are CFT quantities, and the quantities without tildes are those of the AdS black hole)

	\begin{equation}
		\label{2}
		\begin{aligned}
			&\tilde{E}= \frac{M}{\omega}, \qquad 	\tilde{S} = S = \frac{A}{4 G}, \qquad \tilde{T}= \frac{T}{\omega}, \qquad \tilde{\Phi} = \frac{\Phi \sqrt{G}}{\omega\, L},\\
			& \tilde{Q}=\frac{Q L}{\sqrt{G}}, \qquad \tilde{\Omega} = \frac{\Omega}{\omega}, \qquad \text{and}\qquad \tilde{J} = J,
		\end{aligned}
	\end{equation}
	where $S$ and $T$ represent the Bekenstein-Hawking entropy and Hawking temperature, respectively. $\Phi$ and $Q$ denote the electric potential and electric charge, while $\Omega$ and $J$ correspond to the angular velocity and angular momentum. The variation of the mass of black hole $d M$ is written as follow \cite{18}
	\begin{equation}
		\label{1}
		dM = T d\left(\frac{A}{4G}\right) + \Omega dJ + \frac{\Phi}{L} d(QL) 
		+ \left( M - TS - \Omega J - \frac{\Phi}{L} QL \right) \frac{d\left( L^{d-2}/G \right)}{L^{d-2}/G}.	
	\end{equation}
	Moreover, we can express $d\tilde{E}$ as follows
	\begin{equation}
		\label{2ay}
		d\tilde{E}=\frac{d M}{\omega}-\frac{M}{\omega^2} d \omega
	\end{equation}
	Through the AdS/CFT dictionary and Eqs. \eqref{1} and \eqref{2ay}, we build the holographic first law as follows
	\begin{equation}
		\label{7Z}
		d\tilde{E} = \tilde{T} dS + \tilde{\Omega} dJ + \tilde{\Phi} d\tilde{Q} + \mu d C - P dV,
	\end{equation}
	where $C$ is the central charge of CFT, expressed by \cite{28}
	\begin{equation}
		C = \frac{L^{d-2}}{G},
	\end{equation}
	$\mu$ represents the chemical potential, given by
	\begin{equation}
		\mu = \frac{1}{C} \left( \tilde{E} - \tilde{T} S - \tilde{\Omega} J - \tilde{\Phi} \tilde{Q} \right), 
	\end{equation}
	and $P$ is the pressure of  CFT, related to the volume and CFT energy by the following equation of state
	\begin{equation}
		\label{8Z}
		P V = \frac{\tilde{E}}{(d - 2)}.
	\end{equation}
	To eliminate $P dV$ in the first law, we rescale the thermodynamic quantities as follows
	\begin{equation}
		\label{H1}
		\hat{M}=	\hat{E} = \omega L \tilde{E}, \quad \hat{T} = \omega L \tilde{T}, \quad \hat{\Omega} = \omega L \tilde{\Omega}, \quad \hat{\Phi} = \omega L \tilde{\Phi}, \quad \hat{\mu} = \omega L \mu.
	\end{equation}
	By this rescaling, using  Eqs.~\eqref{9Z}, \eqref{7Z}, and \eqref{8Z}, we obtain the first law as follows
	\begin{equation}
		\label{FYY}
		\begin{aligned}
			d\hat{M} &= \omega L\, d \tilde{E} + \tilde{E}\, d(\omega L) \\
			&= \omega L \left( \tilde{T} \, dS + \tilde{\Omega} \, dJ + \tilde{\Phi} \, d\tilde{Q} + \mu \, dC - \frac{\tilde{E} \, dV}{(d - 2)\, V} \right) + \tilde{E}\, d(\omega L).
		\end{aligned}
	\end{equation}
	Finally, we obtain a simplified form of the first law and the Smarr relation as follows
	\begin{equation}
		\label{FL}
		d\hat{M}=\hat{T} dS + \hat{\Omega} dJ + \hat{\Phi} d\tilde{Q} + \hat{\mu} dC,
	\end{equation}
	and
	\begin{equation}
		\label{SR}
		\hat{M}= \hat{T} S + \hat{\Omega} J + \hat{\Phi} \tilde{Q} + \hat{\mu} C.
	\end{equation}
	respectively. By these constructions, we establish a new version of holographic thermodynamics of black holes to solve the issues of other formalisms, where the first law is not degenerate, and the Smarr relation is a homogeneous function of the first order without considering Newton's constant as a variable as  in the restricted phase space thermodynamics.  
	\section{Charged Higher Dimensional AdS Black Holes}
	\label{SS2}
	This section provides a detailed study of the thermodynamic behavior of charged and non-rotating higher-dimensional AdS black holes and their relation to the conformal field theory (CFT) via conformal holographic extended thermodynamics. The charged AdS black hole in \(d\) dimensions is described by the metric \cite{31a}
	\begin{equation}
		\label{m11}
		ds^2 = -f(r)\, dt^2 + \frac{dr^2}{f(r)} + r^2\, d\Omega_{d-2}^2,
	\end{equation}
	where \(d\Omega_{d-2}^2\) denotes the metric on a unit \((d-2)\)-dimensional sphere. The metric function is given by
	\begin{equation}
		f(r) = 1 - \frac{ m}{r^{d-3}} + \frac{q^2}{r^{2(d-3)}} + \frac{r^2}{L^2},
		\	\label{gm}
	\end{equation}
	with \(m\) and \(q\) related to the mass \(M\) and the charge \(Q\) of the black hole via
	\begin{equation}
		m = \frac{16 \pi G M}{(d-2)\, \lambda_{d-2}}, \qquad q = \frac{8 \pi G}{\sqrt{2(d-2)(d-3)} \lambda_{d-2}}\, Q, \qquad \lambda_{d-2} = \frac{2 \pi^{\frac{d-1}{2}}}{\Gamma\left(\frac{d-1}{2}\right)}.
		\label{qt}
	\end{equation}
	Through the Bekenstein–Hawking law, we can express the entropy of a charged black hole in $d$ dimensions as follows
	\begin{equation}
		\label{shg}
		S_{BH} = \frac{\lambda_{d-2}\, r_h^{d-2}}{4G},
	\end{equation}
	where \(r_h\) denotes the radius of the event horizon. To illustrate our purpose, we study two cases of these black holes : the five-dimensional (5D) and six-dimensional (6D) charged AdS black holes.
	
	\subsection{5D Charged AdS Black Holes}
	We explore the holographic thermodynamic behavior and the critical phenomena of the AdS charged black holes in  five dimension spacetime, by using the re-scaling of  CFT dual. Trough  the adaptation of  general metric Eq. \eqref{gm} and Eq. \eqref{qt} where $\lambda_{3}=2 \pi^2$, we express the metric function of five-dimensional charged black holes as follows
	\begin{equation}
		f(r)= 1 - \frac{8GM}{3 \pi r^2} + \frac{4 G^2 Q^2 }{3  \pi^2 r^4} + \frac{r^2}{L^2}.
	\end{equation}
	Also, we  obtain the mass expression in term of the event horizon, $r_h$, and the other parameters by setting $f(r_h)=0$, as follows
	\begin{equation}
		\label{m5a}
		M = \frac{3 \pi r_h^2 \left(L^2 + r_h^2\right)}{8 G L^2} + \frac{G Q^2}{2 \pi r_h^2}.
	\end{equation}
	We use the Bekenstein–Hawking area law, Eq. \eqref{shg}, to express the entropy of charged AdS black holes in five dimensions as 
	\begin{equation}
		S= \frac{\pi^2 r_h^3}{2G}.
		\label{s1}
	\end{equation}
	Also, the central charge of CFT dual is given in term of the AdS radius and the Newton constant as 
	\begin{equation}
		C= \frac{L^3}{G}.
		\label{c1}
	\end{equation}
	To analyze the thermodynamics within the conformal holographic extended thermodynamic  framework, we rewrite the rescaled mass $\hat{M}$ in terms of the extended holographic variables, based on Eqs. \eqref{m5a}-\eqref{c1}
	\begin{equation}
		\hat{M}= \frac{ 6 S^2 + 2 G^5 \pi^2 \tilde{Q}^2 + 3 \cdot 2^{1/3} C^{2/3} G^{4/3} \pi^{4/3} S^{4/3} }
		{4 \cdot 2^{2/3} C^{1/3} G^{8/3} \pi^{5/3} S^{2/3}}.
	\end{equation}
	The first law and the Smarr relation in the framework of the CHET formalism are written respectively as follows
	
	\begin{equation}
		\label{FL1}
		d\hat{M} = \hat{T} \, dS + \hat{\Phi} \, d\tilde{Q} + \hat{\mu} \, dC,
	\end{equation}
	and
	
	\begin{equation}
		\label{SR1}
		\hat{M} = \hat{T} S + \hat{\Phi} \tilde{Q} + \hat{\mu} C.
	\end{equation}
	Based on the first law, all conjugate quantities can be expressed as follows
	\begin{equation}
		\hat{T}=\left(\frac{\partial \hat{M}}{\partial S}\right)_{\tilde{Q}, C}= \frac{ 12 S^2-2 G^5 \pi^2 \tilde{Q}^2 + 3 \cdot 2^{1/3} C^{2/3} G^{4/3} \pi^{4/3} S^{4/3} }
		{6 \cdot 2^{2/3} C^{1/3} G^{8/3} \pi^{5/3} S^{5/3}},
	\end{equation}
	\begin{equation}
		\hat{\Phi}=\left(\frac{\partial \hat{M}}{\partial \tilde{Q}}\right)_{S, C}	= \frac{G^{7/3} \pi^{1/3} \hat{Q}}{2^{2/3} C^{1/3} S^{2/3}},
	\end{equation}
	\begin{equation}
		\hat{\mu}=\left(\frac{\partial \hat{M}}{\partial C}\right)_{S, \bar{Q}}= \frac{-2 G^5 \pi^2 Q^2 + 3 \cdot 2^{1/3} C^{2/3} G^{4/3} \pi^{4/3} S^{4/3} - 6 S^2}{12 \cdot 2^{2/3} C^{4/3} G^{8/3} \pi^{5/3} S^{2/3}}.
	\end{equation}
	We explore the phase structure, the stability  and the impact of CFT dual on the thermodynamics behavior of this black holes in the context of the conformal  holographic extended   thermodynamic. Our analysis begins with the study of critical phenomena, where the critical parameters are determined by solving the following equations
	\begin{equation}
		\left(\frac{\partial \hat{T}}{\partial S}\right)_{\tilde{Q}, C}=0 \quad \text { and } \quad\left(\frac{\partial^2 \hat{T}}{\partial S^2}\right)_{\tilde{Q}, C}=0.
	\end{equation}
	The critical parameters are given by
	\begin{equation}
		S_c = \sqrt{\frac{5}{3}} \, G^{5/2} \pi \tilde{Q}, \qquad C_c  = \frac{6 \sqrt{5\, G} \tilde{ Q}}{\pi}, \qquad \hat{T}_c= \frac{4 \sqrt{3}}{5 \pi  G^2}.
	\end{equation}
	To investigate the phase transition of five-dimensional charged AdS black holes, and the role of their CFT dual, represented by the central charge, in the phase structure, we study the Helmholtz free energy as a function of the rescaled Hawking temperature. Within the CHET formalism, the Helmholtz free energy is given by
	\begin{equation}
		\hat{F} = \hat{M} - \hat{T}\,S= \frac{ -6 S^2 + 3 \cdot 2^{1/3} \, C^{2/3} \, G^{4/3} \, \pi^{4/3} \, S^{4/3} + 10 G^5 \pi^2 \tilde{Q}^2 }{ 12 \cdot 2^{2/3} \, C^{1/3} \, G^{8/3} \, \pi^{5/3} \, S^{2/3} }.
	\end{equation}
	To examine the stability of the black holes phases, we compute the heat capacity \(\zeta\). A positive value of heat capacity indicates stability, while  negative value signals instability. The heat capacity is given by
	\begin{equation}
		\zeta = \hat{T} \left(\frac{\partial S}{\partial \hat{T}}\right)_{\tilde{Q}, C}= \frac{36 \left(\sqrt[3]{2} \pi ^{4/3} C^{2/3} G^{4/3} S^{7/3}+6 S^3\right)}{-15 \sqrt[3]{2} \pi ^{4/3} C^{2/3} G^{4/3} S^{4/3}+50 \pi ^2 G^5 Q^2+60 S^2}-\frac{3 S}{5}
	\end{equation}  
	\begin{figure}[htp]
		\centering
		\includegraphics[width=0.5\linewidth]{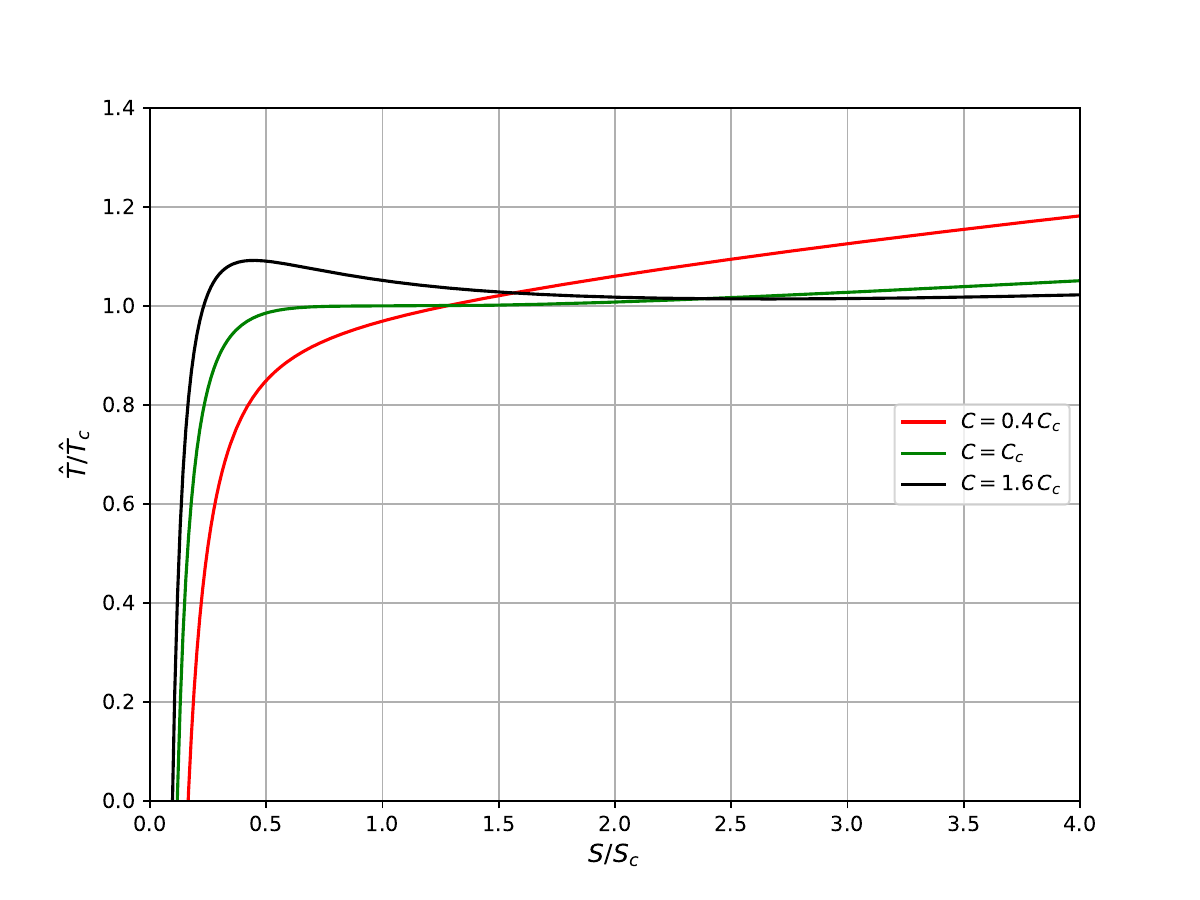}
		\caption{$T-S$ curves of 5D charged AdS black holes for different  states of CFT. }
		\label{ts}
	\end{figure}
	\begin{figure}[htp]
		\centering
		\includegraphics[width=0.5\linewidth]{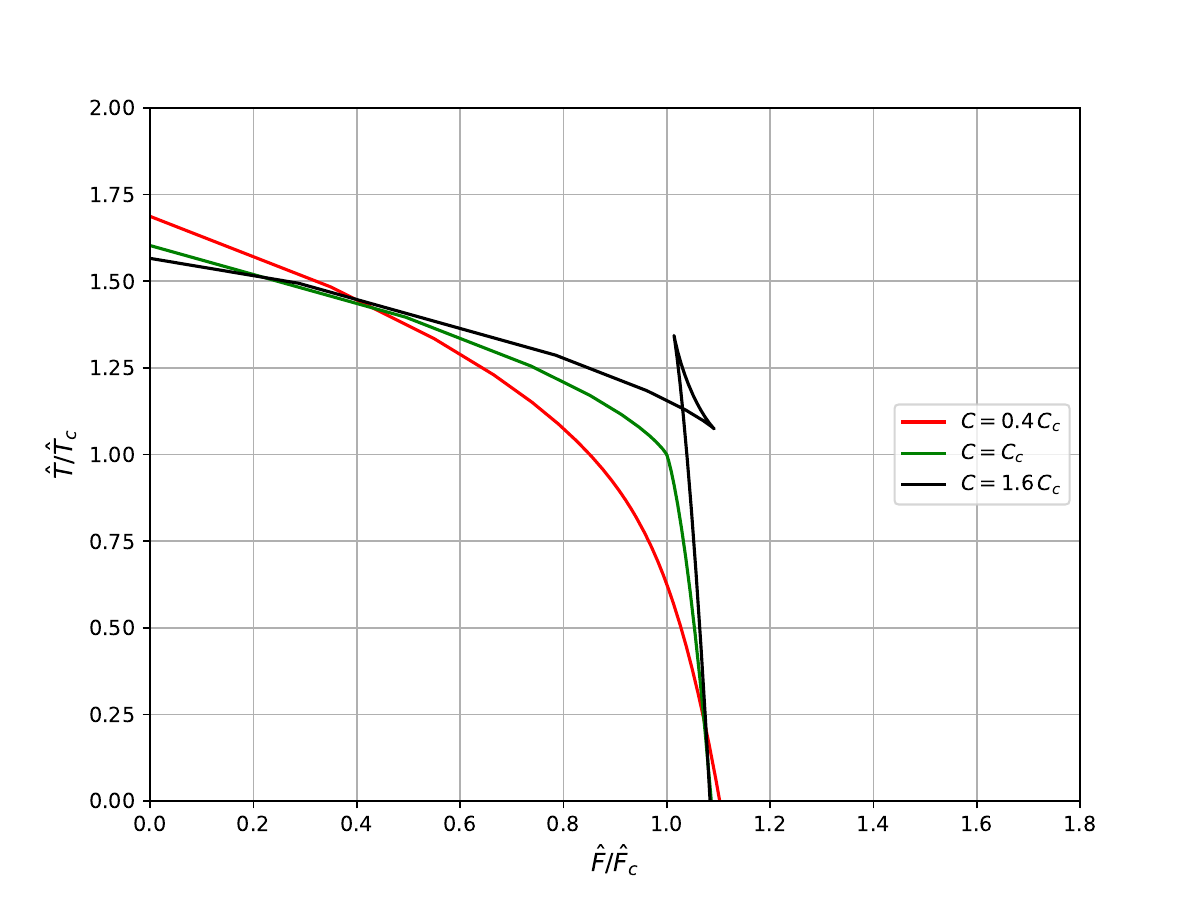}
		\caption{$F-T$ curves of 5D charged AdS black holes for different  states of CFT. }
		\label{ft}
	\end{figure}
	\\
	
	Fig.\ \ref{ts} represents the thermal evolution of charged AdS black holes in  five-dimensional spacetime in terms of the Bekenstein–Hawking entropy for different values of the central charge of  CFT. Fig.\ \ref{ft} illustrates the variation of the Helmholtz free energy as a function of the Hawking temperature for different CFT states. From Figs.\ \ref{ts} and \ref{ft}, we observe three types of thermodynamic behavior of these black holes,  depending on the state of  CFT, and represented by the value of its central charge.
	In the first case, when the central charge of  CFT is greater than the critical value (black curves in Figs.\ \ref{ts} and \ref{ft}), a first-order phase transition occurs between small, medium, and large black holes, where the swallowtail in Fig.\ \ref{ft} represents the first-order phase transition.
	In the second case, when the central charge of  CFT equals the critical value \(C_c\) (green curves in Figs.\ \ref{ts} and \ref{ft}), a second-order phase transition occurs between small and large black holes, with the critical phenomenon at \(S = S_c\).
	In the third case, when the central charge of  CFT is smaller than the critical value (red curves in Figs.\ \ref{ts} and \ref{ft}), no phase transitions occur, and only a single phase is present.
	
	\begin{figure}[htp]
		\centering
		\includegraphics[width=0.4\linewidth]{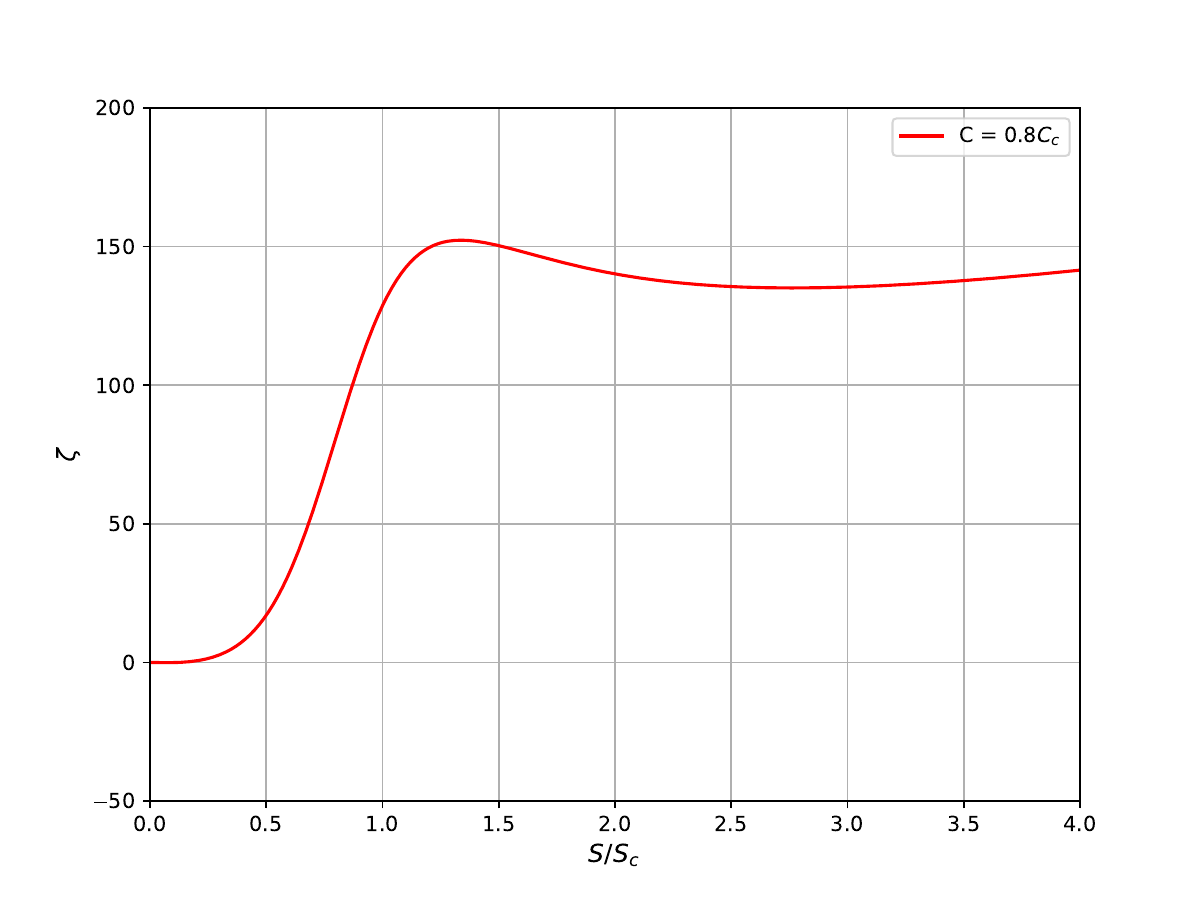}
		\includegraphics[width=0.4\linewidth]{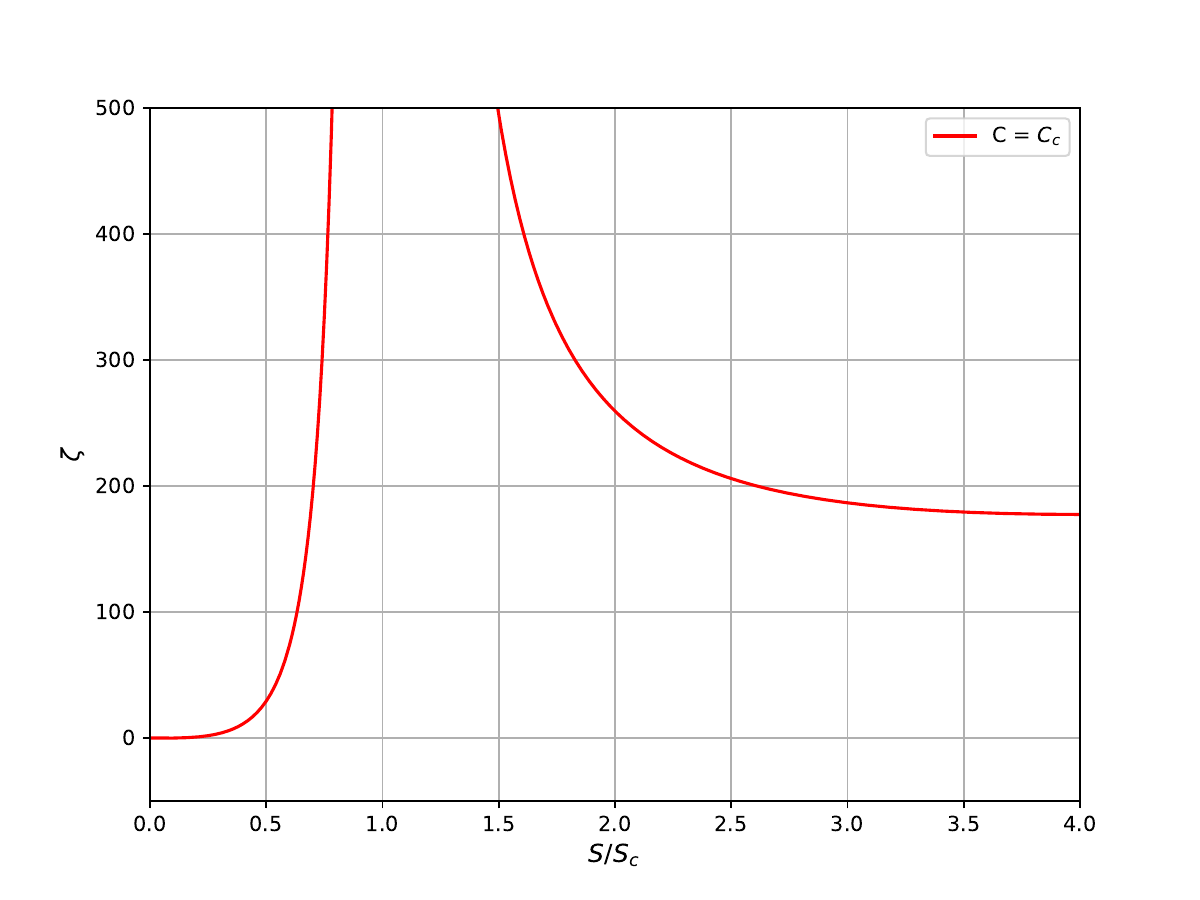}
		\includegraphics[width=0.4\linewidth]{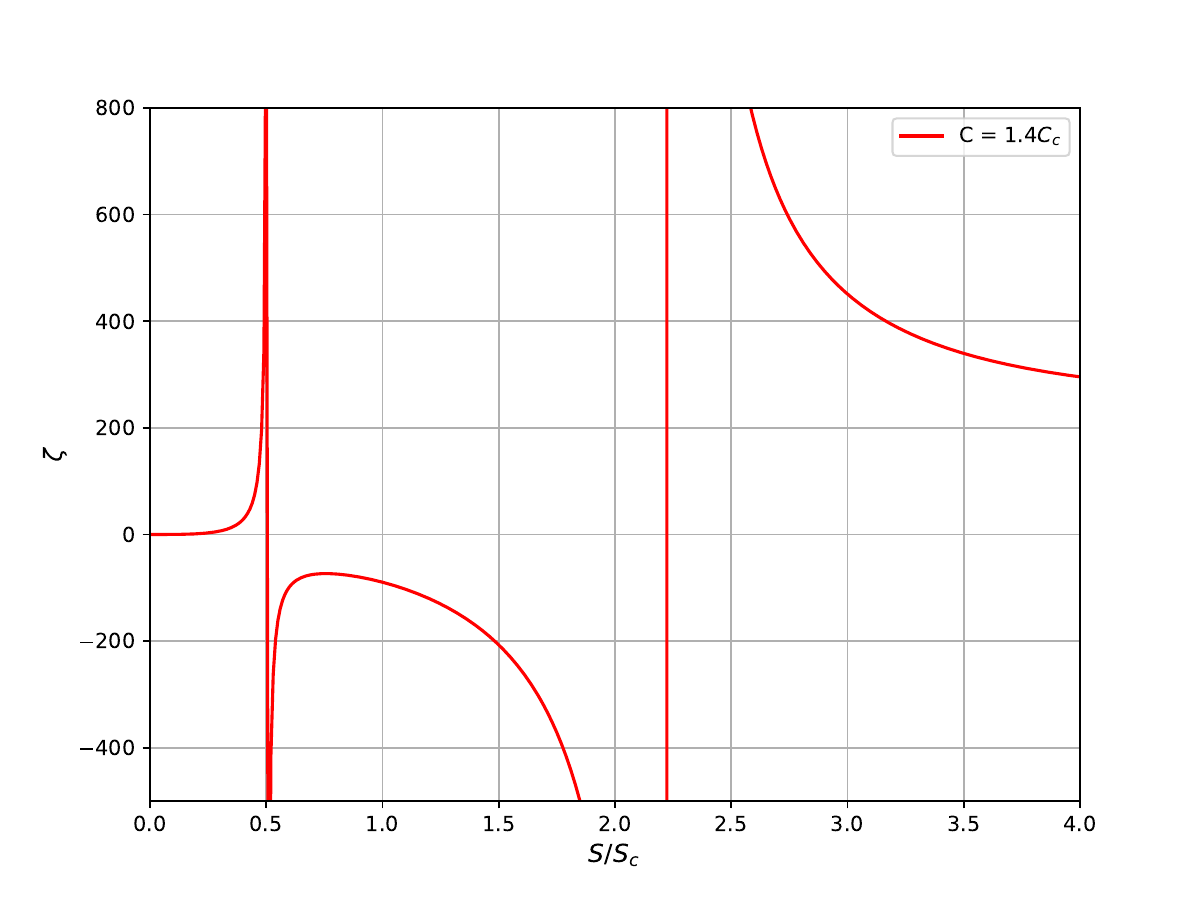}
		\caption{ Heat capacity evolution  of 5D charged AdS black holes for different  states of CFT. }
		\label{hC}
	\end{figure}
	
	Fig.\ \ref{hC} illustrates the heat capacity curves of five-dimensional charged AdS black holes for different values of the central charge of  CFT. These curves reveal the stability and phase structure of the black holes. 
	For \(C > C_c\), a first-order phase transition is represented. In this case, there are three regions corresponding to small, medium, and large black holes. The small and large black holes have positive heat capacity, indicating their thermodynamic stability, while the medium black holes have negative heat capacity, indicating instability.
	For \(C = C_c\), a second-order phase transition occurs. There are two regions corresponding to small and large black holes, both of which exhibit positive heat capacity, indicating they are thermodynamically stable.
	For \(C < C_c\), there is a single region with positive heat capacity, meaning that the black holes are stable and no phase transition occurs in this case. The thermodynamic behavior of black holes within the framework of conformal holographic extended thermodynamics resembles that of a Van der Waals fluid. This behavior is analogous to that observed in the thermodynamics of four-dimensional Reissner–Nordström AdS black holes within this formalism \cite{28}. 
	Through our analysis, we discovered the decisive role of the boundary state in the thermodynamic behavior of gravitational systems in the five-dimensional bulk, as represented by the central charge of  CFT and its corresponding phase structure.  This study demonstrates that the degrees of freedom of the boundary theory, represented by the central charge of  CFT, play a crucial role in determining the behavior of gravitational systems in the bulk.
	\subsection{6D Charged AdS Black Holes}
	After studying the five-dimensional charged AdS black holes, we now turn to the six-dimensional case within the framework of holographic thermodynamics with a rescaled dual CFT. Based on the general higher-dimensional metric given in Eqs.~\eqref{m11}--\eqref{qt}, and using
	$\lambda_{4} = {8\pi^{2}}/{3}$, the metric function takes the form
	\begin{equation}
		f(r) = 1 - \frac{3GM}{2\pi\,r^{3}} + \frac{3G^{2}Q^{2}}{8\pi^{2}\,r^{6}} + \frac{r^{2}}{L^{2}}.
	\end{equation} 
	To study the thermodynamic behavior of six-dimensional charged AdS black holes, we determine the black hole mass by solving the condition \( f(r_h) = 0 \), where \( r_h \) denotes the radius of the event horizon. From this condition, we find that the mass of the black hole is given by 

	\begin{equation}
		\label{m6D}
		M=  \frac{2 \pi  r_h^3 \left(L^2+r_h^2\right)}{3 G L^2} + \frac{G Q^2}{4 \pi  r_h^3}.
	\end{equation}
	From the Bekenstein–Hawking area law, Eq.~\eqref{shg}, the entropy of the six-dimensional charged AdS black hole in terms of the horizon radius is
	\begin{equation}
		\label{SS1a}
		S= \frac{2 \pi^2 r_h^4}{3 G},
	\end{equation}
	and the central charge of the dual CFT is
	\begin{equation}
		\label{c2}
		C= \frac{L^4}{G}.
	\end{equation} 
	We express the rescaled mass of these black holes in terms of CHET quantities, based on the mass formula in Eq.~\eqref{m6D} and using Eqs.~\eqref{SS1a} and \eqref{c2},  as follows
	
	\begin{equation}
		\hat{M}= \frac{ 6 \sqrt{2} S^2+ \sqrt{2} G^5 \pi^2 \tilde{Q}^2 + 4 \sqrt{3} \sqrt{C} G \pi S^{3/2}  }{ 2 \cdot 6^{3/4} C^{1/4} G^{5/2} \pi^{3/2} S^{3/4} }.
	\end{equation}
	To study the thermodynamic behavior and characteristics of the six-dimensional charged AdS black holes in the holographic interpretation, we use the first law of thermodynamics to derive the conjugate thermodynamic quantities as follows
	\begin{equation}
		\hat{T}=\left(\frac{\partial \hat{M}}{\partial S}\right)_{\tilde{Q}, C}= \frac{ 3^{1/4} \left(10 \sqrt{2} S^2 + 4 \sqrt{3} \sqrt{C} G \pi S^{3/2} - \sqrt{2} G^5 \pi^2 \tilde{Q}^2 \right) }{ 8 \cdot 2^{3/4} C^{1/4} G^{5/2} \pi^{3/2} S^{7/4} },
	\end{equation}
	\begin{equation}
		\hat{\Phi}=\left(\frac{\partial \hat{M}}{\partial \tilde{Q}}\right)_{S, C}	= \frac{ G^{5/2} \sqrt{\pi} \tilde{Q} }{ 2^{1/4} \, 3^{3/4} \, C^{1/4} \, S^{3/4} },
	\end{equation}
	and
	\begin{equation}
		\hat{\mu}=\left(\frac{\partial \hat{M}}{\partial C}\right)_{S, \tilde{Q}}= - \frac{ 6 \sqrt{2} S^2 - 4 \sqrt{3} \sqrt{C} G \pi S^{3/2} + \sqrt{2} G^5 \pi^2 \tilde{Q}^2 }{ 8 \cdot 6^{3/4} \, C^{5/4} \, G^{5/2} \, \pi^{3/2} \, S^{3/4} }.
	\end{equation}

	We study the phase transitions, critical phenomena, thermal evolution, and stability of six-dimensional charged AdS black holes, along with the impact of the dual CFT on the thermodynamic behavior. First, we determine the critical point quantities by solving the following equations
	\begin{equation}
		\left(\frac{\partial \hat{T}}{\partial S}\right)_{\tilde{Q}, C}=0, \quad \text{and} \quad\left(\frac{\partial^2 \hat{T}}{\partial S^2}\right)_{\tilde{Q}, C}=0.
	\end{equation}
	The critical quantities are  given by
	\begin{equation}
		S_c = \sqrt{\frac{21}{10}} \, G^{5/2} \pi \tilde{Q}, \qquad C_c=  \frac{20}{9 \pi} \sqrt{\frac{70}{3}} \sqrt{G} \tilde{Q}, \qquad \text{and} \qquad \hat{T}_c= \frac{6 \sqrt{5}}{7 \pi  G^2}.
	\end{equation}
	The Helmholtz free energy writes
	\begin{equation}
		\hat{F}= \hat{M} - \hat{T} S= \frac{ -6 \sqrt{2} S^2 + 4 \sqrt{3} \sqrt{C} G \pi S^{3/2} + 7 \sqrt{2} G^5 \pi^2 \tilde{Q}^2 }{ 8 \cdot 6^{3/4}  C^{1/4} \, G^{5/2}  \pi^{3/2}  S^{3/4} }.
	\end{equation}
	The heat capacity of these black holes is expressed as
	\begin{equation}
		\zeta=\hat{T} \left(\frac{\partial S}{\partial \hat{T}}\right)_{\tilde{Q}, C}= \frac{ 4 S \left( 10 \sqrt{2} S^2 + 4 \sqrt{3} \sqrt{C} G \pi S^{3/2} - \sqrt{2} G^5 \pi^2 \tilde{Q}^2 \right) }{ 7 \sqrt{2} G^5 \pi^2 \tilde{Q}^2 - 4 \sqrt{3} \sqrt{C} G \pi S^{3/2} + 10 \sqrt{2} S^2 }.
	\end{equation}
	
	\begin{figure}[htp]
		\centering
		\includegraphics[width=0.5\linewidth]{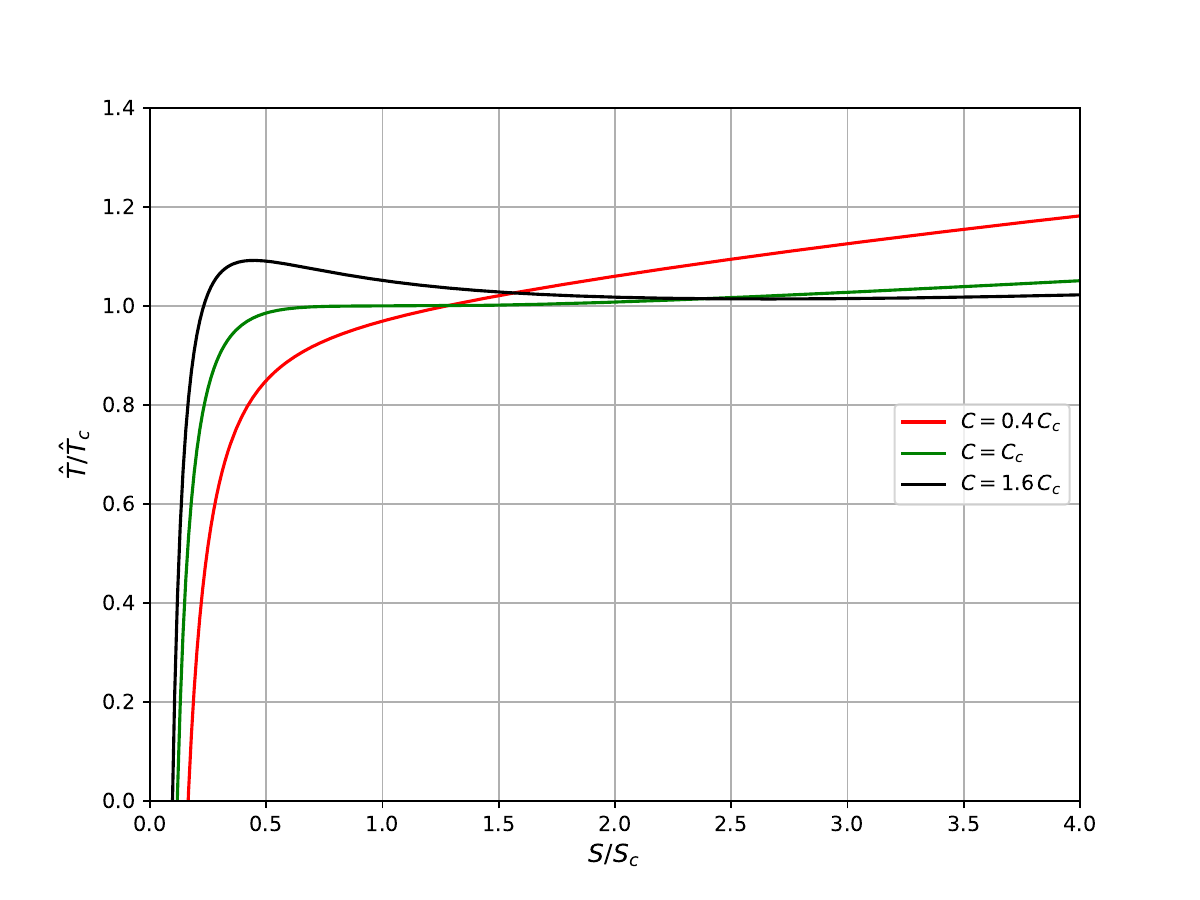}
		\caption{$T$–$S$ curves of 6D charged AdS black holes for different states of  CFT.}
		\label{ts6}
	\end{figure}
	
	\begin{figure}[htp]
		\centering
		\includegraphics[width=0.5\linewidth]{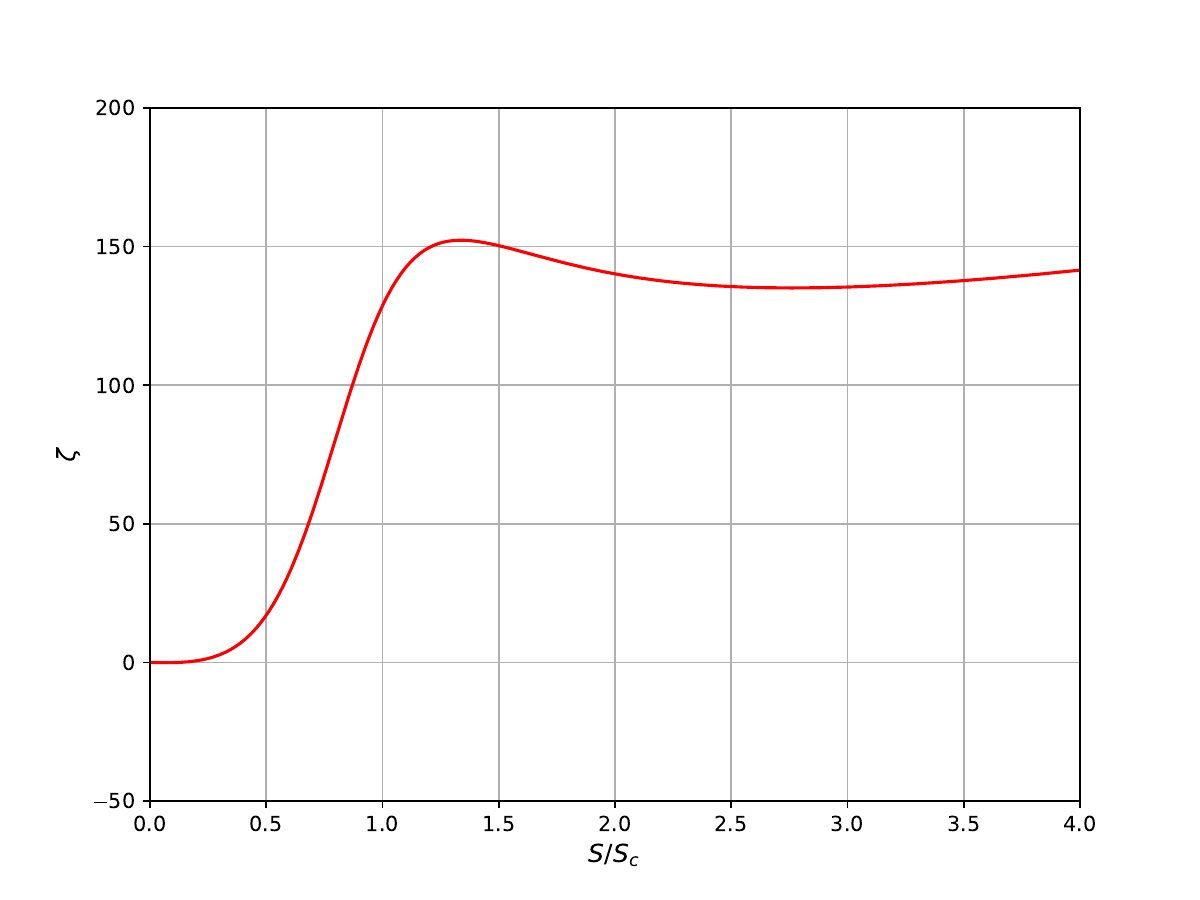}
		\caption{$F$–$T$ curves of 6D charged AdS black holes for different states of CFT.}
		\label{ft6}
	\end{figure}
	
	\begin{figure}[htp]
		\centering
		\includegraphics[width=0.4\linewidth]{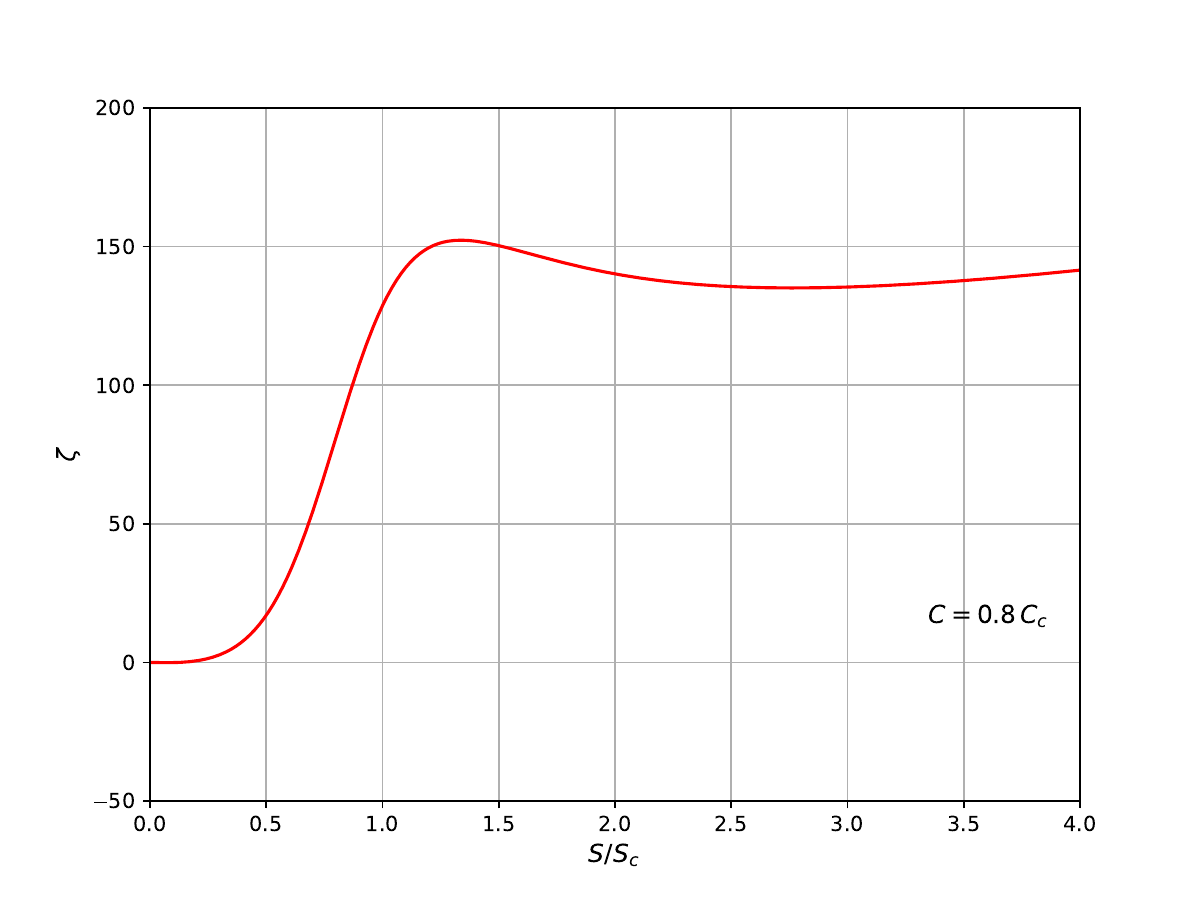}
		\includegraphics[width=0.4\linewidth]{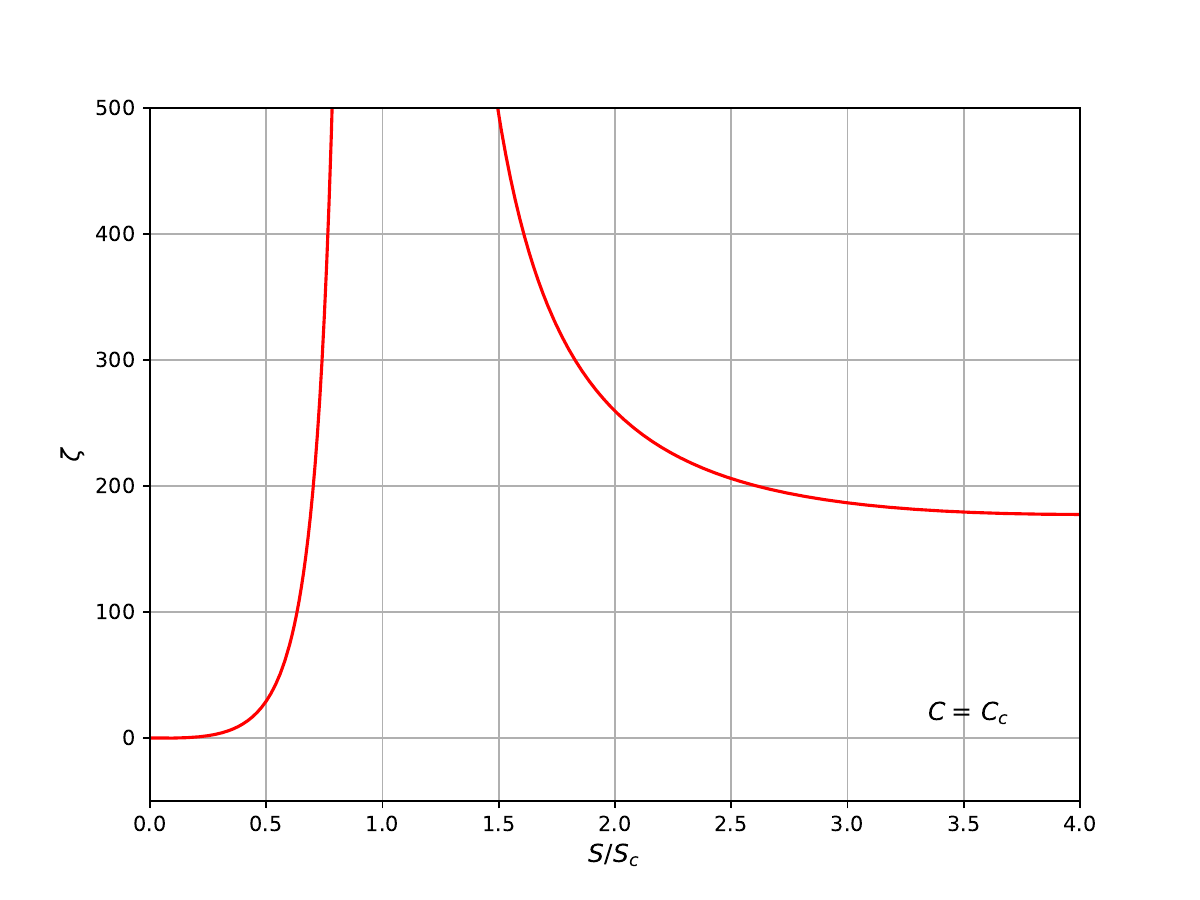}
		\includegraphics[width=0.4\linewidth]{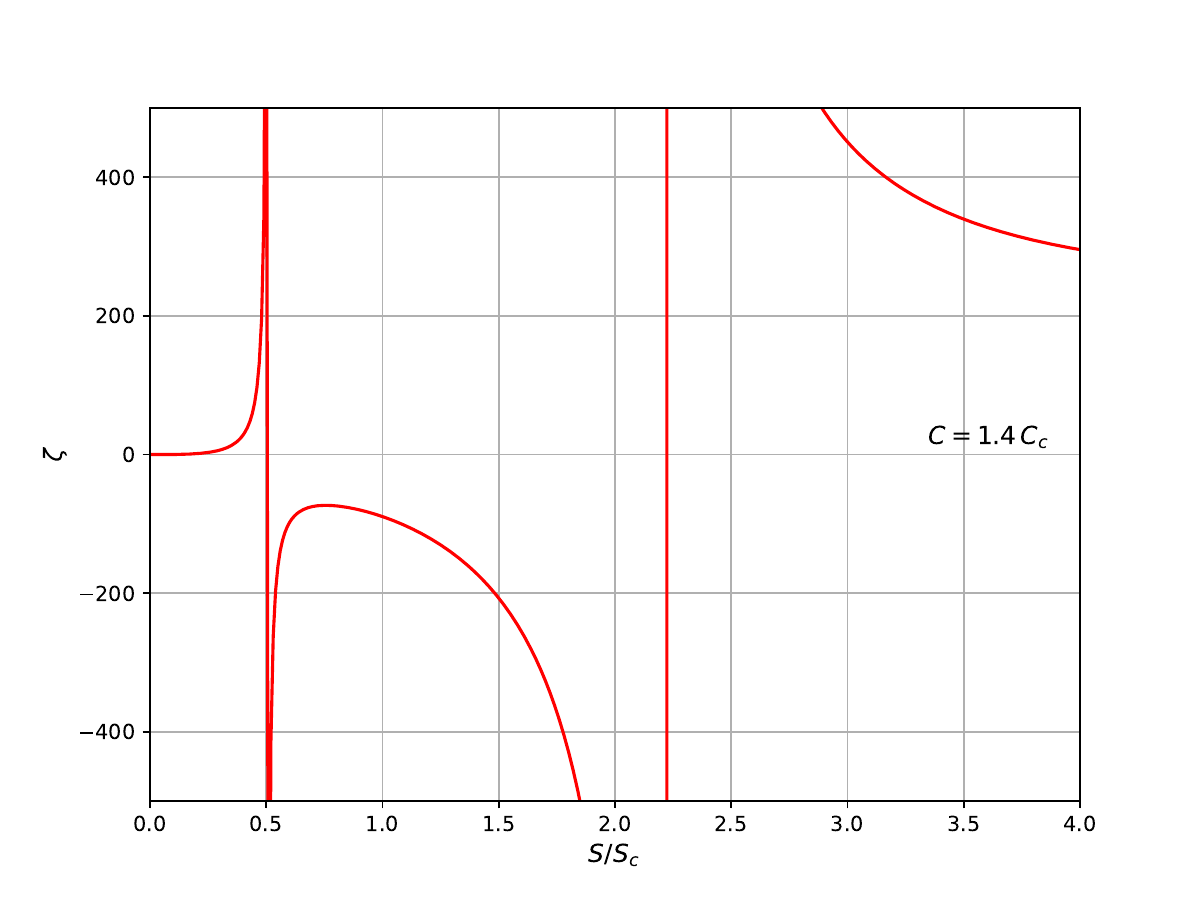}
		\caption{Heat capacity evolution of 6D charged AdS black holes for different  states of CFT.}
		\label{hC6}
	\end{figure}
	Through Figs. \ref{ts6}–\ref{hC6}, we investigate the phase structure of charged AdS black holes in six-dimensional spacetime and examine the role of the boundary (CFT) on the thermodynamic behavior of the bulk (AdS black holes) within the conformal holographic extended thermodynamics framework. In Fig. \ref{ts6}, we depict the thermal evolution in terms of the black hole entropy. Fig. \ref{ft6} illustrates the Helmholtz free energy versus temperature; and Fig. \ref{hC6} shows the heat capacity as a function of entropy. To highlight the influence of the boundary on the thermodynamic behavior of the bulk, we plot the evolution of these quantities for all states of  CFT, which are characterized by the central charge \(C\). There are three states, each corresponding to a specific range of \(C\) and exhibiting distinct thermodynamic behavior and phase structure for the six-dimensional charged AdS black holes.
	For the first state, \(C > C_c\) (black curves in Figs. \ref{ts6}–\ref{hC6}), a first-order phase transition occurs among small, medium, and large black holes, as indicated by the swallowtail in the Helmholtz free energy curve. Small and large black holes have positive heat capacities (thermodynamically stable), whereas medium black holes have negative heat capacities (unstable). In the second state, \(C = C_c\) (green curves), a second-order phase transition takes place between small and large black holes at the critical entropy \(S = S_c\). Both small and large black holes exhibit positive heat capacities, indicating stability. In the third state, \(C < C_c\) (red curves), only a single black hole phase exists, with positive heat capacity and no phase transition.
	\\
	
	We conclude from these results that the thermodynamic behavior and phase structure of six-dimensional charged AdS black holes within the conformal holographic extended thermodynamics framework closely resemble those of their four- and five-dimensional counterparts~\cite{28}, and they are also similar to the phase behavior of a Van der Waals fluid. In addition, the thermodynamic behavior of six-dimensional black holes is directly influenced by the degrees of freedom of the boundary CFT. These findings highlight how the state of the boundary theory governs the evolution of gravitational systems in the bulk. They also provide a foundation for understanding the quantum origin of black hole phase transitions in higher dimensions, which cannot be observed in lower-dimensional systems. For example, BTZ black holes, which represent charged AdS black hole solutions in three dimensions, exhibit no critical phenomena or phase transitions and remain thermodynamically stable for all values of the central charge \cite{30}.  Therefore, higher-dimensional black holes within the AdS/CFT correspondence offer a promising framework for investigating the origin of black hole phase transitions, thermodynamic behavior and the evolution of gravitational systems in the bulk.

	\section{Conclusion and Discussion}
	\label{SS3}
	
	We studied higher-dimensional charged AdS black holes in the context of the AdS/CFT correspondence, using the rescaling of the CFT while respecting conformal symmetry, and treated the rescaling factor of the CFT as a thermodynamic variable. In this approach, Newton’s constant is held fixed while the AdS radius is allowed to vary. This framework is referred to as conformal holographic extended thermodynamics.
	\\
	
	Our focus has been on the influence of the boundary theory on the thermodynamic behavior of black holes in the bulk. Specifically, we investigated the impact of the central charge of the boundary theory on the thermal evolution of the bulk. The central charge is interpreted as a counting parameter for the boundary degrees of freedom. In the AdS/CFT correspondence, $C \propto N^2$ for SU($N$) gauge theory, so varying the central charge corresponds to moving between different boundary theories. From the bulk perspective, the variation of $C$ in the boundary corresponds to changes in the on-shell action and the partition function, which in turn leads to variations in the thermodynamic properties of the bulk.
	\\

	We analyzed the thermodynamic behavior and phase structure of charged AdS black holes in five and six dimensions across all possible values of the central charge. For large central charge, specifically when \(C > C_c\), we found the existence of a first-order phase transition between small, medium, and large black holes. In this regime, the small and large black holes are thermodynamically stable, while the medium black holes are unstable. At the critical value of the central charge, \(C = C_c\), the system undergoes a second-order phase transition between small and large black holes. When the central charge is small, \(C < C_c\), we observed that no phase transition occurs and the system remains in a single stable phase.  
	\\
	
	These results resemble the well-known Van der Waals-like behavior exhibited by four-dimensional Reissner–Nordström AdS black holes~\cite{28}. However, they contrast sharply with the three-dimensional BTZ black hole case, where no phase transitions or critical phenomena occur for any value of the central charge~\cite{30}. This highlights the significant role played by spacetime dimensionality. While higher dimensions, \(D \geq 4\), exhibit nontrivial thermodynamic phenomena including criticality and phase transitions, lower-dimensional spacetimes lack such complex thermodynamic structure.    
	\\
	
	In summary, conformal holographic extended thermodynamics provides a compelling framework for understanding how boundary degrees of freedom influence thermodynamics in the bulk. Future investigations may extend this study to include rotating black holes, thereby offering deeper insights into the rich interplay between boundary and bulk thermodynamics.

	\section*{Acknowledgments}
	Y. Ladghami gratefully acknowledges the support from the "PhD-Associate Scholarship – PASS" grant provided by the National Center for Scientific and Technical Research in Morocco, under grant number 42 UMP2023.
	
\end{document}